\documentclass{aa}  

\usepackage{graphicx}
\usepackage{txfonts}
\begin{document} 

   \title{Improving CME evolution and arrival predictions with AMR and grid stretching in Icarus}

   %\subtitle{?}

   \author{T. Baratashvili \inst{1}, C. Verbeke \inst{1, 2}, N. Wijsen \inst{1},
           S. Poedts \inst{1,3}
          }

   \institute{Department of Mathematics/Centre for mathematical Plasma Astrophysics, 
             KU Leuven, Celestijnenlaan 200 B, 3001 Leuven, Belgium,
             \email{tinatin.baratashvili@kuleuven.be}
             \and
             Royal Observatory of Belgium, Ringlaan 3, 1180 UKKel, Belgium
             \and
             Institute of Physics, University of Maria Curie-Sk{\l}odowska, 
             PL-20-031 Lublin, Poland}

\date{Received: \today}
\titlerunning{CME evolution and arrival predictions with AMR and grid stretching}
\authorrunning{Baratashvili et al.}
% \abstract{}{}{}{}{} 
% 5 {} token are mandatory
 
  \abstract
  % aims headTVDLFing (mandatory)
   {Coronal mass ejections (CMEs) are one of the main drivers of disturbances in interplanetary space. Strong CMEs, when directed towards the Earth, cause geomagnetic storms upon interacting with the Earth's magnetic field, and can cause significant damage to our planet and affect everyday life. As such, efficient space weather prediction tools are necessary to forecast the arrival and impact of CME eruptions. Recently, a new heliospheric model called Icarus was developed based on MPI-AMRVAC, which is a 3D ideal magnetohydrodynamics (MHD) model for the solar wind and CME propagation, and it introduces advanced numerical  techniques to make the simulations more efficient. In this model the reference frame is chosen to be co-rotating with the Sun, and radial grid stretching together with adaptive mesh refinement (AMR) can be applied to the numerical domain. }
   %Aim
   {Grid stretching and AMR speed up simulation results and performance. Our aim is to combine the advanced techniques available in the Icarus model in order to obtain better results with fewer computational resources than with the equidistant grid. Different AMR strategies are suggested, depending on the purpose of the simulation.  }
   {In this study, we model the CME event that occurred on July 12, 2012. A cone model was used to study the CME's evolution through the background solar wind, and its arrival at and impact with the Earth. 
   %For maximum efficiency and speed-up, 
   Grid stretching and AMR were combined in the simulations by using multiple refinement criteria, to assess its influence on the simulations' accuracy and the required computational resources. We compare simulation results to the EUHFORIA model.}
  % Results
   {We applied different refinement criteria to investigate the potential of solution AMR for different applications. As a result, the simulations were sped up by a factor of $\sim$17 for the most optimal configuration in Icarus. For the cone CME model, we found that limiting the AMR to the region around the CME-driven shock yields the best results. The results modelled by the simulations with radial grid stretching and AMR level 4 are similar to the results provided by the original EUHFORIA and Icarus simulations with the `standard' resolution and equidistant grids. The simulations with 5 AMR levels yielded better results than the simulations with an equidistant grid and standard resolution. }
   {Solution AMR is flexible and provides  the user the freedom to modify and locally increase the grid resolution according to the purpose of the simulation. We find that simulations with a combination of grid stretching and AMR can reproduce the simulations performed on equidistant grids significantly faster. 
   The advanced techniques implemented in Icarus can be further used to improve the forecasting procedures, since the reduced simulation time is essential to make physics-based forecasts less computationally expensive. 
   }


   \maketitle
%
%-------------------------------------------------------------------

\section{Introduction} \label{section:introduction}
Conditions in the solar environment, particularly in the vicinity of the Earth, are extremely important for human life on our planet, which is ever more dependent on advanced technology.
Coronal mass ejections (CMEs) are one of the main drivers of space weather \citep{Gopalswamy2017}. They are violent eruptive events, during which large clouds of plasma (up to $10^{16}\;$g) are released from the solar corona at speeds ranging from $100$ to $3000\;$km s$^{-1}$, based on SOHO/LASCO measurements, with an average of about $450\;$km s$^{-1}$ \citep{Webb2006}. 
The occurrence of CMEs strongly correlates with the magnetic cycles of the Sun. During the solar minimum, CMEs occur about once every five days, whereas during the solar maximum, two or more CMEs can be observed per day \citep{Park2012}. 

Coronal mass ejections with speeds (in the local reference frame with respect to the Sun) higher than the local fast magnetosonic speed, generate shocks that proceed them as they travel across the heliosphere. In fact, even relatively slow CMEs with an initial velocity that is only slightly higher than the velocity of the background solar corona, often develop a shock wave further away from the Sun, typically beyond 50 solar radii. The fast magnetosonic frequency in the ambient plasma density decreases rapidly and the CME, subsequently, becomes super-sonic and super-fast \citep{Talpeanu2022}.

Currently, the advanced space weather forecasting tools that are operationally used are physics-based models. One such recently developed prediction tool is the EUropean Heliospheric FORecasting Information Asset (EUHFORIA; \citet{Pomoell2018}). This heliospheric wind and CME propagation and evolution model is used both for scientific research and for forecasting purposes. The very similar model ENLIL \citep{Odstrcil2004} is even more widely used for space weather forecasting purposes. Both these prediction tools  relax the 3D ideal magnetohydrodynamics (MHD) equations in the heliosphere in order to obtain a steady background solar wind model.
In both of these simulation tools, CMEs can be superposed on the steady background wind. They both include a so-called cone CME model \citep{Zhao2002, Xie2004, Xue2005}. This CME model simulates a CME by approximating it as a spherical hydrodynamic plasma blob that is inserted at the inner boundary of the simulation domain (0.1~AU). Hence, this CME model aims to simulate the CME shocks and enables a forecast of their arrival times at the Earth, other planets, and satellites. It does not contain any of the CME's magnetic features. 

There are  alternative models that start from much closer to the Sun and include more physics, such as the Alfv\'en-Wave driven Solar wind Model  \citep[AWSoM;][]{Sokolov2013,Toth2012}. This model is more advanced and starts at the chromosphere. Within the AWSoM model, grid stretching is included to resolve the transition region, as well as adaptive mesh refinement (AMR) to increase the resolution near the heliospheric current sheet. However, the inclusion of the physics source terms, and in particular the extension of the numerical domain to the chromosphere, comes at a price. In spite of the advanced numerical techniques, AWSoM requires too much CPU time to be useful in an operational forecasting environment. Therefore, a faster version is currently being developed that starts a bit higher up in the solar atmosphere to relax the time-step limitation: AWSoM-R \citep{Vanderholst2014,Jin2017}. 

A new heliospheric wind and CME propagation model, called Icarus \citep{Verbeke2022}, has recently been developed and implemented within the EUHFORIA2.0 project. The Icarus model has been implemented within the  MPI-AMRVAC framework \citep{Xia2018}. MPI-AMRVAC is a parallel AMR framework that is well suited for solving systems of (near-)conservation laws, such as the equations of hydrodynamics and MHD. Unlike EUHFORIA, which uses Heliocentric Earth EQuatorial (HEEQ) coordinates, the new Icarus model uses a reference frame that is co-rotating with the Sun and it assumes time-independent solar wind boundary conditions in this frame. As a result, it yields a truly steady-state solar wind configuration in the computational domain once the MHD relaxation phase is over. 

The novelty of Icarus, however, lies in applying advanced numerical techniques to the computational domain, enabling a substantial speed-up of the computations. \citet{Verbeke2022} presented the Icarus model and briefly discussed the effect of radial grid stretching and solution AMR applied separately to the simulations, to determine how each of these technical features affects the duration of the simulation and its accuracy. The authors reported that both radial grid stretching and solution AMR can speed up the simulations notably, while still producing results similar to equidistant simulations. The speed-up is important since the wall-clock time, the actual time taken from the start of the simulation to the end, is important in an operation space weather forecasting context, because predictions are made ahead of time to enable mitigation of space weather damages. Moreover, reliable predictions are not based on a single simulation: the uncertainty on the input parameters can be quite substantial \citep{Riley2018}. These uncertainties can be taken into account by so-called `ensemble' modelling techniques, which have become an important technique in the past decade to improve the reliability of space weather forecasts \citep{Guerra2020}. In space weather, single-model ensembles are most common, consisting of a number of simulation runs with the same model but with perturbations of the initial conditions (see e.g.  \citet{Mays2015} for an example of CME propagation ensemble modelling). Since CME arrival time models have a large number of CME input parameters, for instance, ten for a simple spheromak CME model \citep{Verbeke2019}, such ensembles can become large, and require significant computational time. Hence, the `need for speed'.

In the present paper, we further investigate the capabilities and potential of the new heliospheric model Icarus. We investigated if and how the combination of solution AMR with radial grid stretching can be used to substantially decrease the duration of the simulations but yet obtain the same accuracy. We introduce specific AMR criteria that were selected for different purposes. Alternatively, Icarus can obtain better and higher-resolution results (for instance with better resolved CME shock fronts) in the same wall-clock time as the current equidistant Icarus and EUHFORIA simulation runs. As explained by \cite{Xia2018} and \cite{Verbeke2022}, radial grid stretching produces grid cells that become larger with increasing distance from the inner boundary. Even though the resolution of the grid decreases with increasing radial distances from the Sun, the AMR feature is used to increase the local resolution, wherever it is needed or wanted. In this paper, we demonstrate the flexibility of AMR and discuss the use of different AMR criteria, depending on the purpose of the simulation. The considered criteria are focused on refining the CME interior (when studying CME deformation or erosion, for instance), the CME shocks (e.g.\ when studying shock arrival times or when using the shock information as input for a solar energetic particle evolution and transport model, such as PARADISE \citep{Wijsen-PhD}), or the combination of both. 

For this purpose, we simulated the CME event of July 12, 2012, which has already been investigated in detail by \citet{Scolini2019} using the EUHFORIA model. 
In order to assess the effect of the combination of grid stretching and the different AMR criteria, we compare the results obtained with the advanced techniques in Icarus to the results produced by the simulations with an equidistant grid without AMR.

The paper is organised as follows: the Icarus model settings, the event specifications, and the modelling parameters  are described in Sect.~\ref{section:tools}; the different AMR criteria are illustrated in Sect \ref{section:amr_section}; and the results are discussed and conclusions presented in Sect.~\ref{section:discussion_conclusion}.

\section{Simulation methods and setup} \label{section:tools}
\subsection{Icarus} \label{tools_icarus}

Icarus is a new heliospheric model for the solar wind and interplanetary CME propagation that was recently developed by \citet{Verbeke2022}. The model is implemented into MPI-AMRVAC, which is a parallel AMR framework, solving partial differential equations with various numerical schemes \citep{Xia2018}.
In order to model the solar heliosphere, the set of ideal MHD equations is solved in a frame that is co-rotating with the Sun. In this model, we used a Total Vanishing Diminishing Lax-Friedrichs (TVDLF) scheme, which is second-order accurate both in time and space. Since we are dealing with shocks in our simulation, we used a limiter for the numerical scheme. We performed a study involving many different combinations of schemes and limiters for Icarus, and we selected the most optimal combination, namely the TVDLF scheme with a Woodward limiter \citep{Woodward1984}. This combination yields the best accuracy for the least wall-clock time. We applied parabolic cleaning in order to minimise $\nabla \cdot \mathbf{B}\,$ errors.  The computational domain covers the heliocentric radial distances from $0.1\;$AU to $2\;$AU, including the orbit of the inner planets and Mars, 360 degrees in the longitudinal direction and 120 degrees ($\pm 60^{\circ}$) in the latitudinal direction, excluding the poles. Since the inner boundary conditions for the solar wind are time-independent, after the relaxation phase has passed, the modelled solar wind converges to a time-independent steady-state solution in the given reference frame that is co-rotating with the Sun. The relaxation period is determined by the time that is required for the slow solar wind stream to traverse the computational domain from the inner boundary towards the outer boundary.
Once the relaxed solar wind solution is obtained, CMEs can be injected from the inner heliospheric boundary. In this work, we introduced the CME into the domain by varying the inner radial boundary conditions using a cone CME model. The cone model can have various implementations and an overview can be found in \cite{Scolini2018}. 
Here, the implementation corresponding to run~6 of \cite{Scolini2018} was chosen, which is also applied in \cite{Verbeke2022}. This implementation describes the geometry of a sphere passing through a spherical boundary. The radius of the CME was kept constant as it passed through the inner heliospheric boundary and the half radius $r_{1/2}$ upon the interaction with the inner heliosphere is defined as follows: 
\begin{align} \label{realistic_radius}
    r_{1/2} = R \cdot \text{sin} \;  \Bigg(\frac{\omega_{\text{CME}}}{2}\Bigg),
\end{align}
where $R$ is the distance of the inner heliospheric boundary $0.1\;$AU (21.5 $R_\odot$) and $\omega_{CME}$ is the full width of the CME. Another important parameter to characterise the CME is the opening angle, which is the angle that the CME subtends at the intersection with the inner heliospheric boundary. The opening angle is given by
\begin{align} \label{spherical}
    \theta(t) =\text{ arccos} \Bigg (\frac{R_c(t)^2 + R^2 - r_{1/2}^2}{2 \cdot R_c(t)\cdot R}\Bigg),
\end{align}
where $R_c(t)$ is the distance of the CME centre from the solar centre and it is given as follows: 
\begin{align} \label{rc_equations}
R_c(t) = t \cdot v_{\text{CME}} + R - r_{1/2},
\end{align}
where $v_{\text{CME}}$ is the CME injection speed. 

Finally, in order to identify where the CME is present at the inner heliospheric boundary of the domain, we determined all points that satisfy the condition $d \leq \theta(t)$, where $d$ is the angular distance from any point $p$ on the inner heliospheric domain to the centre of the CME insertion, and is given by
\begin{align}
\begin{split}
    d &= \arccos\bigg(\sin\theta_p \sin \theta_{CME}  \cos(\phi_p - \phi_{CME}) +\\& + \cos (\theta_p) \mbox{cos}(\theta_{CME})\bigg),
\end{split}
\end{align}
where ($\theta_{CME}, \phi_{CME}$) corresponds to the co-latitude and longitude of the launch direction of the CME, and $(\theta_p, \phi_p)$ corresponds to the co-latitude and longitude of a certain point $p$ on the inner boundary shell.

The solar wind is considered supersonic at the inner heliospheric boundary and we can set the plasma conditions there from the data provided by any coronal model that is in agreement with this assumption. For the numerical scheme that we used, two layers of ghost cells were introduced at the inner and outer boundaries of the domain. We prescribed the plasma variables provided by the coronal model at the inner boundary 21.5$\;R_\odot$. The boundary conditions were assumed to be constant in time in the reference frame co-rotating with the Sun and we ignored the differential rotation of the Sun. More details on the boundary conditions utilised in Icarus can be found in \cite{Verbeke2022}.
In this study, we  used the coronal model implemented in EUHFORIA. This model uses the semi-empirical
Wang–Sheeley–Arge model \citep[WSA;][]{Arge2004} to
generate the solar wind plasma and magnetic quantities at
0.1 au \citep[see][for more details]{Pomoell2018}.
In all the results presented in this paper, the boundary conditions for the solar wind were set the same way for the simulations performed with EUHFORIA and with Icarus. 

\subsection{Grid stretching} \label{subsection:grid_stretching}
In order to model the solar wind at the Earth's location and beyond, a large radial distance needs to be covered. In the Icarus model, as well as in EUHFORIA, the inner and outer radial boundaries of the domain are at $0.1\;$AU and $2\;$AU, respectively. 

In EUHFORIA and the original Icarus model, the mesh is equidistant in all three spherical coordinates, which leads to the broadening of the cells closer to the outer boundary. 
This is because the angular and the radial widths of the cells remain constant, such that the width of the cells (i.e. $ r \Delta \theta$, where $\Delta \theta$ is the width of the cell) increase closer to the outer boundary. 
As a result, the cells there become elongated compared to the cells near the inner heliospheric boundary.

One of the important advantages of the Icarus model is grid stretching in the domain. 
In Icarus, stretching is applied only in the radial direction. The technique is described in \cite{Xia2018} and it maintains the aspect ratio of the width and the length of the cell in the whole domain. In Table~\ref{tab:resolutions} we compare the number of cells in the domain for each resolution for the stretched and equidistant grids. The number of cells in the radial direction was chosen with the formula given in \cite{Xia2018}, that is, 
\begin{equation} \label{number_of_cells}
    N = \frac{\log(\frac{r_{out}}{r_{in}})}{\log(\frac{2+\Delta \theta}{2- \Delta\theta})},
\end{equation}
where $r_{in}$ and $r_{out}$ are the inner and outer radii of the domain, and $\Delta\theta$ is the angular resolution in radians in the longitudinal direction. For the given configuration of the domain, the optimal number of cells in the radial direction is 46 in the low-resolution configuration, obtained from the equation above, but taking into consideration the optimal number of cells and blocks in MPI-AMRVAC \citep{Verbeke2022}, the best result was modelled by choosing $N = 60$ in the radial direction for the low-resolution simulations.
Further details of the applied grid stretching in this model are given in \cite{Verbeke2022}. In this work, the same type of grid stretching was applied. In order to demonstrate the effect of grid stretching in the domain, the meshes are shown in Fig.~3 in \cite{Verbeke2022}. This figure shows that the cells in the domain with grid stretching maintain a cubical shape throughout the domain. 
\begin{figure}[hbt!]
    \centering
    \includegraphics[width=0.5\textwidth]{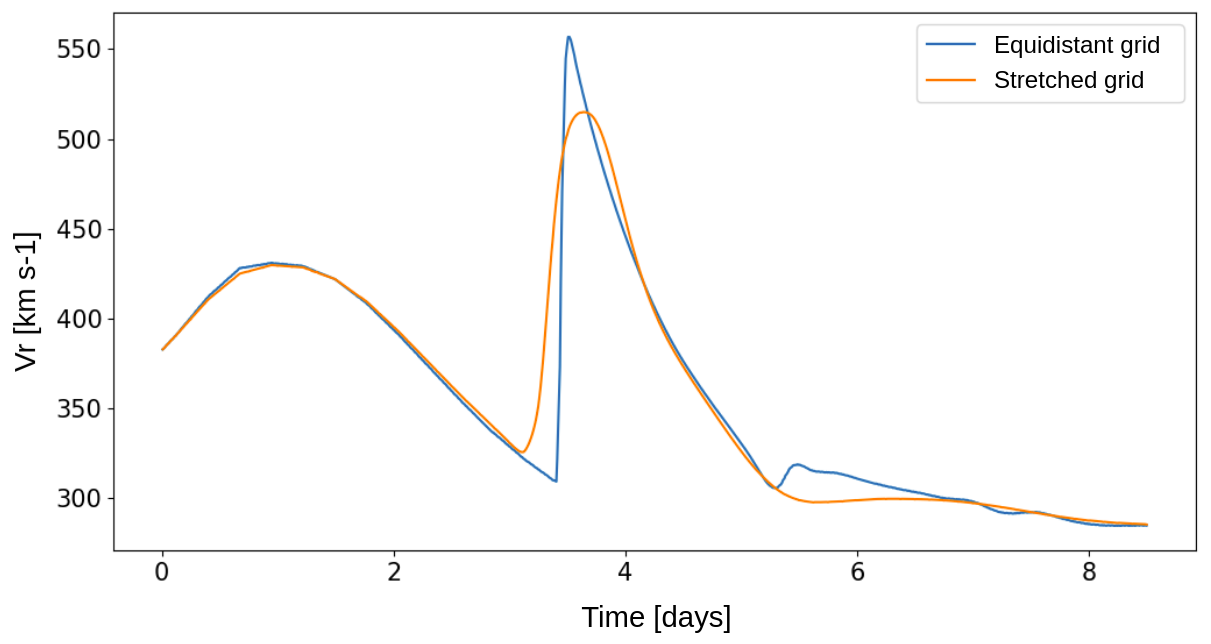}
    \caption{Time series at the Earth of the equidistant (blue) and stretched (orange) low-resolution (given in Table~\ref{tab:resolutions}) grid simulations. The horizontal axis shows the time elapsed from the start of the simulation in days and the vertical axis shows the radial velocity values.}
    \label{fig:str_vs_nonstr}
\end{figure}

A drawback of grid stretching is that the resolution of the domain in the radial direction decreases towards the outer boundary. This typically results in grid cells that are larger at the location of the Earth than in an equidistant grid with the same resolution simulations. For example, in the low-resolution domain (see Table~\ref{tab:resolutions}), where the radial grid stretching is not applied and the cells have equidistant radial lengths, the cell size in the radial direction is $1.362\;R_\odot$ everywhere in the domain for the low-resolution case, that is, with 300 radial grid points (so the radial grid size of 1.9~au/300). In contrast, the radial width of the cells in the stretched grid is  $10.7\;R_\odot$ at 1~AU  for the low-resolution domain. 
The radial length of the cell in the middle resolution for the stretched grid with the same stretching factor is $5.36\;R_\odot$ and for the high-resolution domain it is $2.68\;R_\odot$. On the other hand, the radial resolutions for the equidistant middle- and high-resolution grids are 0.68~$R_\odot$ and $0.34\;R_\odot$, respectively. In this study we only consider the middle-resolution equidistant simulations with Icarus and EUHFORIA as a reference, since this resolution is chosen as the operational resolution in EUHFORIA.

The difference in the domain resolution becomes important when, for example, a CME is propagating in the heliosphere, forming shocks upon interaction with the solar wind.
\begin{table} [hbt!]
\begin{tabular}{|l|l| l|}
\hline
Resolution & \multicolumn{2}{|c|}{\# cells[$r,\theta,\phi$]}
\\\hline\hline
 & Non-stretched & Radially stretched \\\hline
Low& [300,\;32,\;96]& [60,\;32,\;96]\\\hline
Middle& [600,\;64,\;192]& [120,\;64,\;192]\\\hline
High& [1200,\;128,\;384]& [240,\;128,\;384]\\\hline
\end{tabular}
\caption{List of equidistant (non-stretched) and radially stretched grid resolutions referred to throughout this paper.\label{tab:resolutions}}
\end{table}

The effect of grid stretching can be seen while comparing the time series at 1~AU. Two simulations with low-resolution grids are compared in Fig.~\ref{fig:str_vs_nonstr}, but one was performed on an equidistant and one on a stretched grid. It is seen that in this case, the CME-driven shock is less sharp for the stretched grid simulation, due to the lower radial resolution at 1~AU. 

Despite the fact that grid stretching makes the simulations faster, the radial resolution at the locations of interest needs to be improved, but only locally if possible, in order to not increase the use of CPU time unnecessarily. 
This can be achieved by combining the radial grid stretching with AMR. 

\subsection{Event description} \label{section:event_description}
In this work, we study a well-observed CME event with good remote observations of the Sun and clear in situ measurements at the Earth during the CME impact. The CME occurred on July 12, 2012 from NOAA AR 11520. This active region first appeared on the July 5, 2012 on the eastern part of the solar disk \citep{Shen2014}. The CME followed a strong X1.4 class flare, which occurred at 15:37~UT on July 12, 2012 and was observed first by the LASCO C2 coronagraph at 16:48 UT. The observed projected linear speed was 885 km s$^{-1}$ \citep{Scolini2019}. The event was  studied by several other authors, including \cite{Hu2016} and \cite{Gopalswamy2018}. \citet{Scolini2019} modelled this CME event with EUHFORIA. 
In this study, we used the same parameters and equations to introduce the cone CME in the simulations. 
The cone model parameters are summarised in  Table~\ref{table:cme_parameters}, where $t_0$ is the relaxation time in the simulation,  $\theta_{CME}$ and $\phi_{CME}$ are the CME injection coordinates in HEEQ coordinates, $v_{CME}$ is the speed of the CME, and $\omega_{CME}/2$ is the half-width of the CME.

\begin{table}[h!]
\caption{Input parameters of the cone CME model used in this study. The CME was launched after 14 days of background wind relaxation.
}
 \centering
  \begin{tabular}{|c|c|}
    \hline
    Parameter & Value \\
    \hline\hline
     $t_0$& 14 [days]  \\ \hline
     $\theta_{CME}$& -8 $^\circ$  \\ \hline
    $\phi_{CME}$& -4 $^\circ$   \\ \hline
     $v_{CME}$ & 1266 [km s$^{-1}$] \\ \hline
     $\omega_{CME}/2$ & 38$^\circ$\\\hline
    \end{tabular}
 \label{table:cme_parameters}
\end{table}

\section{Solution AMR Strategies}\label{section:amr_section}
The solution AMR is a powerful feature of the Icarus model. More information on the block-adaptive implementation of AMR in MPI-AMRVAC can be found in \cite{keppens2003}. In MPI-AMRVAC, the solution AMR is triggered by prescribing refinement criteria that depend on the local plasma properties (i.e.\ on the time-dependent solution). 
In the regions of the computational domain where the (time-dependent) solution satisfies these criteria, the resolution of the grid is increased. As soon as the criteria are no longer satisfied, the grid is coarsened again. 
Each additional refinement level in the domain splits the cells in two, in all three coordinate directions. The refinement is introduced gradually on the computational grid, with a difference of only one refinement level between the adjacent blocks of grid cells. 
The size of the blocks can be chosen and the refinement conditions can be diverse. Moreover, the user can choose the maximum number of refinement levels to be implemented to obtain the solution. Therefore, this type of AMR gives freedom to `modify' the resolution of the grid locally, depending on the purpose of the simulation. 
Thus, AMR can provide higher resolution in the domain only at specific areas, which results in saved computational time and computer resources.

As mentioned in Sect. \ref{subsection:grid_stretching}, the resolution at the location of the Earth is much lower than the resolution closer to the Sun when only radial grid stretching is applied in the domain. 
If we want to resolve the shock features along the Sun-Earth line better, it is necessary to have a higher resolution at these areas in the domain, which can be achieved using AMR.

In all the Icarus simulations with AMR presented in this paper, the base refinement level was chosen to be the low resolution described in Table~\ref{tab:resolutions}. Hence, AMR level 2 and 3 simulations have a resolution that corresponds, in the most refined areas of the grid, to the middle and high resolutions listed in  Table~\ref{tab:resolutions}, respectively.
Table~\ref{table:amr_resolutions_at_l1} summarises the resulting resolutions at the first Lagrange point L1 for different AMR levels combined with a stretched grid in the presented simulations.
In the following sections, when comparing the results of different AMR simulations with the original equidistant EUHFORIA and Icarus simulations, we use the middle-resolution equidistant grid simulation results for the latter two cases. 
For both cases, the resolution in the domain is uniform. 
The radial resolution for EUHFORIA simulation is 0.802~R$_\odot$ and for Icarus 0.685~R$_\odot$.
The radial resolution at L1 in the case of the AMR level 5 simulation is 0.672~R$_\odot$, which is 1.2 and 1.02 times smaller than in the equidistant EUHFORIA and Icarus simulations, respectively.

\begin{table}[htb!]
\caption{Resolutions at L1 for different simulations with the stretched grid. No AMR corresponds to the low-resolution simulation.}   
\label{table:amr_resolutions_at_l1}   
\centering            
\begin{tabular}{c c c c c}         
\hline\hline  
No AMR & AMR 2 & AMR 3 & AMR 4 & AMR 5 \\ 
\hline            
    10.7 R$_\odot$ & 5.36R$_\odot$ & 2.68 R$_\odot$ & 1.344R$_\odot$  & 0.672 R$_\odot$ \\
\hline                                             

\end{tabular}
\end{table}

\begin{figure}
    \centering
    \includegraphics[width=0.5\textwidth]{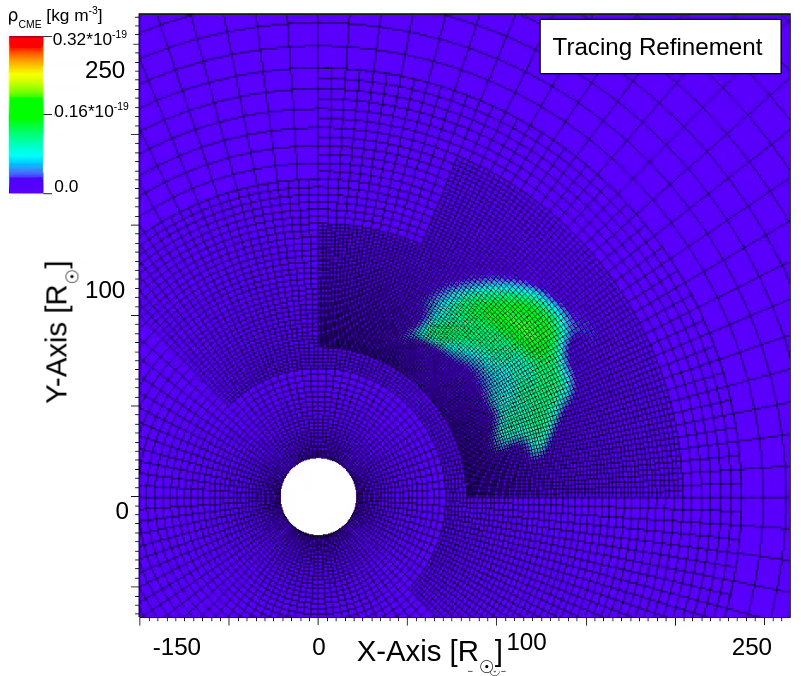}
    \caption{Adaptive mesh refinement level 3 for the CME density tracing function. The $X$- and $Y$-axis show the distance from the centre in solar radii. The CME density tracing function is shown in the equatorial plane with the corresponding colour map.}
    \label{fig:amr_level_3_tracing_function}
\end{figure}

\begin{figure*}[hbt!]
    \centering
    \includegraphics[width=\textwidth]{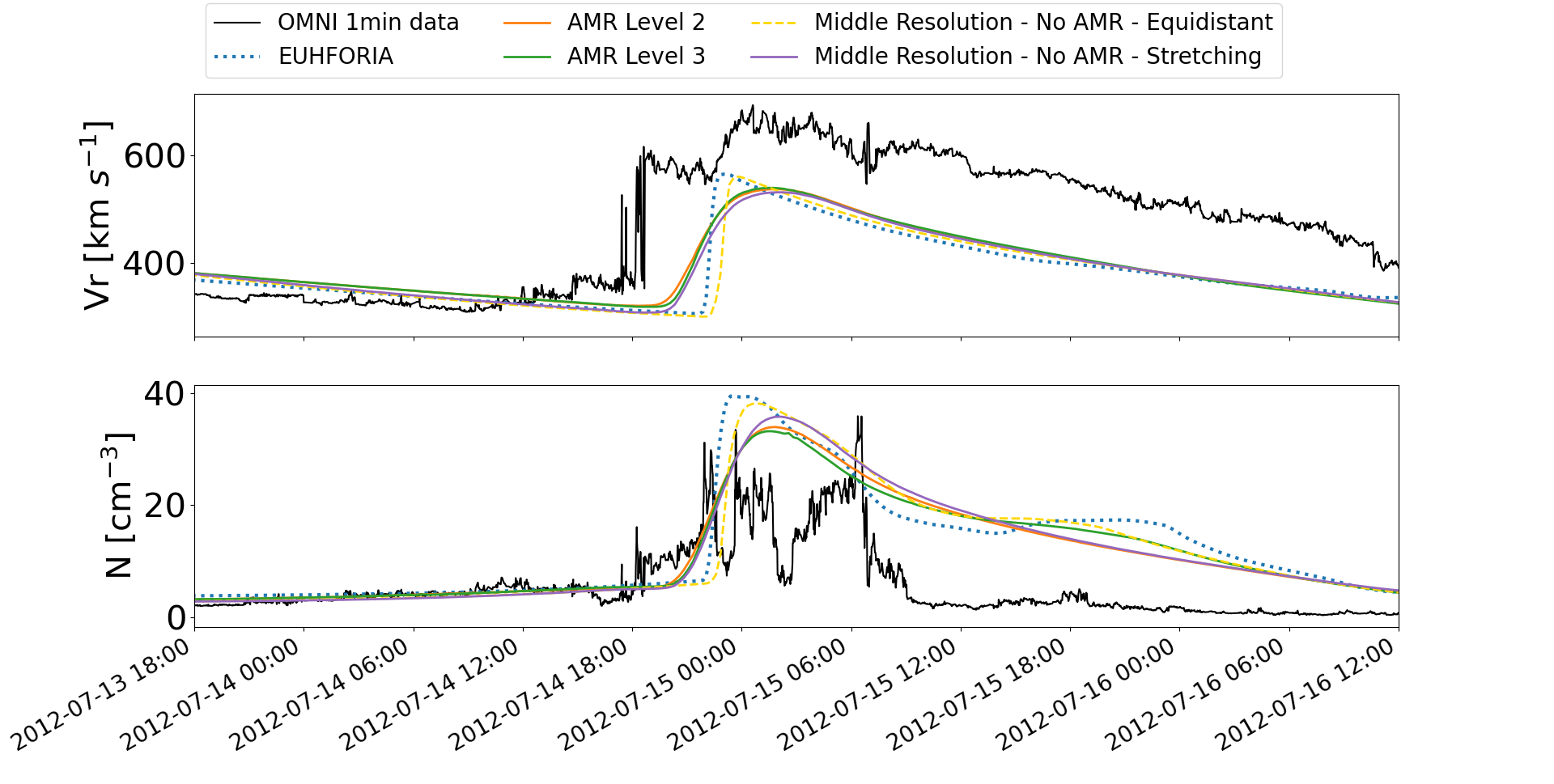}
    \caption{Time series at the Earth for the radial velocity values given in km s$^{-1}$ and number density given in cm$^{-3}$. The tracing function is used for the AMR criteria. The black curve shows the observed values, the blue curve shows the original EUHFORIA result for the high-resolution simulation, and the orange, green, yellow, and purple curves correspond, respectively, to the AMR level 2 simulation,  the AMR level 3 simulation, the middle-resolution simulation without AMR performed on an equidistant grid without stretching, and the middle-resolution run without AMR with grid stretching applied to the computational domain. The horizontal axis shows the time evolution.}
    \label{fig:obs_vs_amr_tracing_euhforia}
\end{figure*}

Adaptive mesh refinement  is only efficient and useful when an appropriate refinement criterion is found. For every different case, depending on the purpose of the simulation, the AMR criterion and the corresponding thresholds should be specified such that only in those parts of the domain where the solution requires a high(er) resolution, the grid is sufficiently refined. \citet{Verbeke2022} already showed that the AMR criterion can be set such that an entire co-rotating interaction region (CIR) in the background wind solution or the CME volume is refined in a non-stretched grid. Here we consider the combination of radial grid stretching and AMR.

In order to investigate CME propagation, including\ its evolution, erosion, and deformation in the heliospheric domain, it is important to have high resolution of the entire CME volume as it propagates through the heliosphere. 
Apart from the CME locations, it is interesting to obtain a high spatial resolution at the CME-driven shock wave. 
Moreover, in some cases when we inject strong CMEs, which drive shock waves, it is necessary to use a combined criterion and refine both the volume of the CME and the CME-driven shock. In this study, we focus on: 
(i)~the refinement of the entire CME volume as it propagates in the heliosphere (Sect. \ref{section:dynamic_refinement}); 
(ii)~the refinement of only the shock produced by the CME interacting with the heliospheric wind (Sect.~\ref{section:shock_refinement}); and 
(iii)~the combination of both (i) and (ii) (Sect.~\ref{section:combined_AMR}).
These three refinement approaches are considered in the following subsections.

\subsection{CME density tracing refinement} \label{section:dynamic_refinement}
The first AMR criterion that we consider follows the entire CME plasma through the interplanetary medium. 
This is achieved by injecting a density tracer together with the CME into the simulation domain. 
The simulation grid is refined wherever and whenever the value of this density tracer function is non-zero, in other words,\ in whatever volume the CME plasma evolves. 
We note that the CME-driven shock and the sheath behind it are formed before the CME itself, namely in regions that consist of solar wind plasma, and thus in which the density tracer function is zero. As a result, the shock and sheath are not necessarily refined, unless they are located very close to the CME, so for very small shock stand-off distances. 

\begin{figure}[hbt!]
    \centering
    \includegraphics[width=0.5\textwidth]{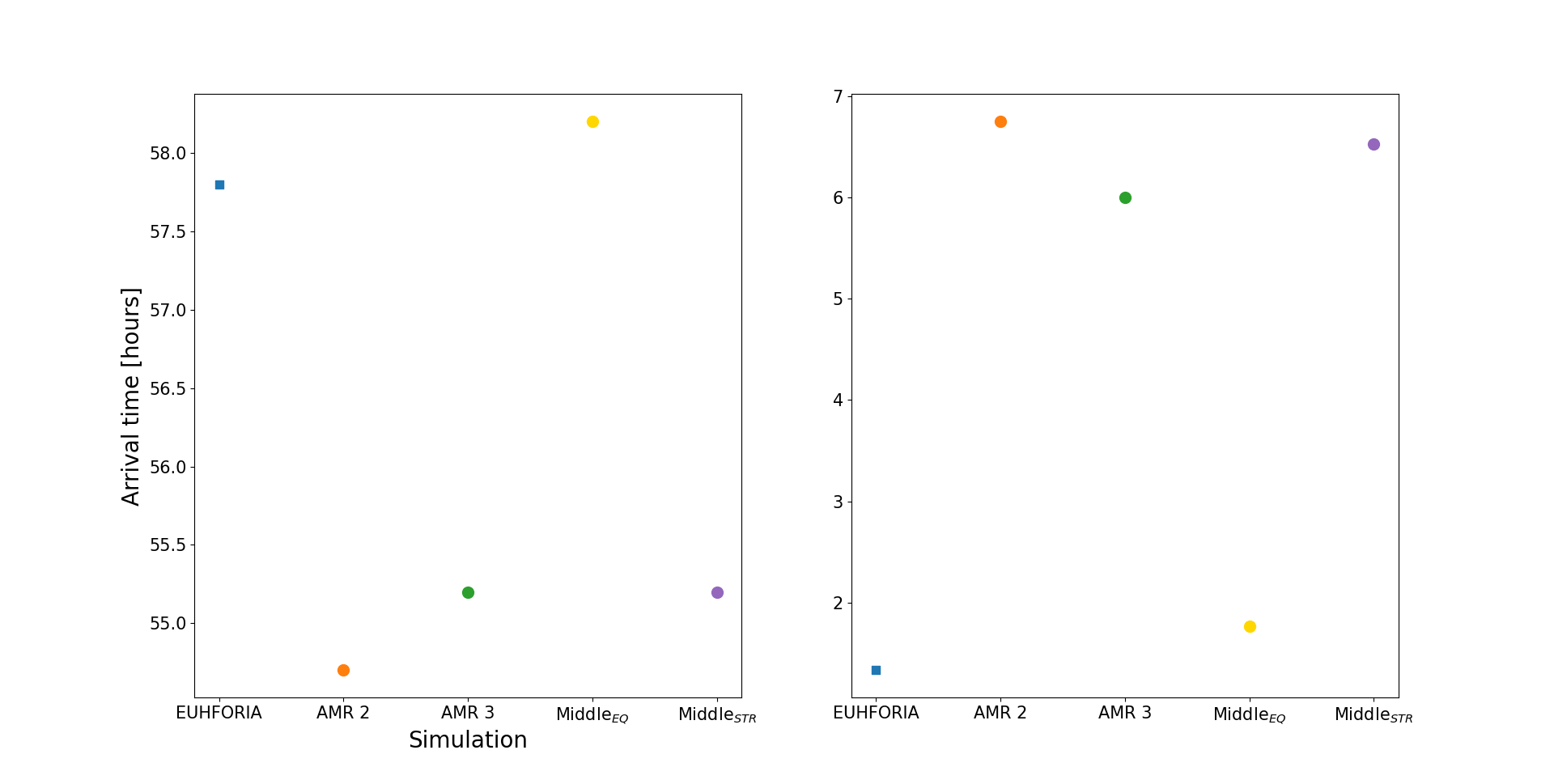}
    \caption{Arrival times (left) and shock widths (right) upon arrival for the original EUHFORIA simulation, AMR level 2 and level 3 runs, and the middle-resolution runs on the equidistant and stretched grids. The vertical axis indicates the simulation type and the horizontal axis shows the arrival times and shock widths in $\;h$. }
    \label{fig:tracing_arrival_shock_width}
\end{figure}

Figure~\ref{fig:amr_level_3_tracing_function} shows the density tracer function together with the underlying computational grid. 
It can be seen that, where the density tracer value is non-zero, the grid is refined to level 3 (set as the maximum refinement level for this simulation). 
This region is surrounded by the refinement level 2 blocks and the rest of the domain is not refined.  
When the CME has passed a certain region, the grid in that region is coarsened again, avoiding unnecessary refinements. As such, we can see that the areas near the inner heliospheric boundary, behind the CME, are coarsened to the base refinement level.

Figure~\ref{fig:obs_vs_amr_tracing_euhforia} shows the time-series data for the radial velocity (upper panel) and number density (bottom panel) at the location of the Earth. 
The black curve represents OMNI-1min data and the blue curve corresponds to the EUHFORIA simulation, which is performed on the middle-resolution domain with equidistant grid spacing. 
The other curves represent the equidistant and the different AMR results obtained with the Icarus model. 

The CME arrives $\sim\;$6 hs later in the simulations than in the OMNI-1min data. The mean absolute errors on the CME arrival time forecasts are around 10$\;h$ \citep{Riley2018}. The shock sharpness and strength are modelled relatively well in the highest-resolution, equidistant simulations. Since, for the chosen set of parameters, the CME arrival occurs later than in the observations, we can see that in the lower-resolution simulations, the time of arrival is closer to the arrival time in the observed data than in the high-resolution simulations. This does not indicate that these simulations model the CME arrival best, as this occurs due to the fact that the shock is more diffuse and smeared-out in the low-resolution simulation. In the simulations, the radial velocity at L1 is considerably lower after the CME has passed, compared to the observations. We believe that this is mostly due to the simplicity of the cone CME model that is used here, and which does not capture the magnetic cloud of the CME. The uncertainties from the modelled solar wind and the uncertainties on the CME parameters also contribute to the discrepancy between the observed and the modelled data. The modelled data show that the cases using an equidistant grid mimic the observed data best. 

In Figure~\ref{fig:obs_vs_amr_tracing_euhforia} we can also see that the EUHFORIA and middle-resolution no-AMR equidistant (Middle$_{EQ}$) curves are most similar. This is because these results are obtained with very similar settings for the computational domain of the EUHFORIA and Icarus models. The EUHFORIA model turns out to capture the shock slightly better than Icarus, in spite of the slightly smaller radial mesh size of Icarus at L1 (0.802~R$_\odot$ for EUHFORIA and 0.685~R$_\odot$ for Icarus). The shock strength, however, is the same for both simulations. The differences in the numerical schemes and computational methods used in EUHFORIA and Icarus are explained in \cite{Verbeke2022}.

It is notable that the three radial velocity profiles of the different simulations using a stretched grid and AMR are quite similar. 
As mentioned above, this is because the utilised AMR criterion refines only the CME, and not the CME-driven shock and its sheath. 
In addition, the cone CME model utilised in these simulations does not contain any internal structure upon its insertion in the computational domain; the initial density is uniform in the CME. As a result, we do not see any significant improvement in the density time-profile at L1 when using higher AMR levels either. 
The CME refinement criterion is expected to be more interesting for CME models with a more complex internal structure, such as\ a Linear Force-Free Spheromak model \citep{Verbeke2019}.

In order to further analyse the results, we calculated the arrival time of the CME shock and the shock width upon arrival for each simulation. The left panel of Fig.~\ref{fig:tracing_arrival_shock_width} shows the arrival times for all the simulations shown in Fig.~\ref{fig:obs_vs_amr_tracing_euhforia} using the same colours. 
This figure shows that the arrival times for the original EUHFORIA simulation and middle-resolution no-AMR equidistant (Middle$_{EQ}$) simulation are very similar, yet they are different from the arrival times for the other simulations. 
Indeed, the shock arrival times for the equidistant simulations with EUHFORIA and Icarus correspond to 57.8 and 58.2~h, respectively.
The shocks arrive earlier in the simulations performed on stretched grids. The shock in the AMR level 2 simulation arrives the earliest, namely at 54.7~h. This is 3.1~h earlier than in the EUHFORIA simulation.
We can also see that the arrival times for the middle-resolution no-AMR stretching (Middle$_{STR}$) simulation and AMR level 3 simulation (also with grid stretching) are very similar, 54.97 and 54.99~h, respectively.
Hence, better resolving the CME plasma does not affect the arrival time of the CME shock.

On the right panel of Fig. \ref{fig:tracing_arrival_shock_width}, we plotted the shock widths for the same simulations. 
A measure of the shock widths was obtained by calculating how long it takes to reach the maximum speed (after the speed jump) after the shock arrival time. It is therefore given in $\;h$. 
The narrowest shocks, lasting for $\sim 1.3\;$h and $\sim 1.7\;$h, were modelled by the equidistant EUHFORIA and Icarus Middle$_{EQ}$ runs, respectively. This is because in these simulations, the radial grid sizes at L1 are the smallest.
The shock widths are much larger for the simulations with stretched grids, and only slightly better for the AMR level 3 simulation since the shock is not better resolved with the applied AMR criterion. 
We can notice that the shock widths are most similar and the largest for AMR level 2 (starting form the Low$_{STR}$ grid) and Middle$_{STR}$ simulations, because locally, at the refined areas the resolution is the same and the shock is equally badly resolved in both simulations. 

From these simulations, we can conclude that the CME density tracing condition for the refinement criterion does not improve the results significantly, because this criterion is not appropriate for capturing the shock and the shock sheath. In this study, we assess the AMR refinement criteria by its ability to model the shock and the CME arrival time accurately, since the cone CME model does not have a complex internal structure. The density tracing criterion only refines the CME interior to the higher resolution, which thus has little effect for a cone CME. On the other hand, the density tracing criterion does not target the shock or the shock sheath, as these regions consist of piled up shocked solar wind in front of the CME. The density tracing criterion will, however, be very useful for quantifying the deformation of a CME and the plasma magnetic flux erosion due to its interaction with the background solar wind when we consider magnetised CME models. Therefore, the density tracing refinement criterion will be mainly used for more advanced flux-rope models, with complex structures and an internal magnetic field. 
In order to study the effect of AMR on the shock, we need to implement a criterion that is specifically aimed at shocks within the domain, which is presented in the next section.

\begin{table}[htb!]

  \caption{Run times required for the EUHFORIA simulation, middle-resolution simulations on the equidistant (Middle$_{EQ}$) and stretched (Middle$_{STR}$) grids, and AMR level 2 and 3 runs (starting from Low$_{STR}$). All the simulations were performed on 1 node with 36 processors on the Genius cluster at the Vlaams Supercomputing Centre.}
  \centering
   \begin{tabular}{c c c c c }
  \hline\hline
   EUHFORIA & Middle$_{EQ}$ &  Middle$_{STR}$ & AMR2& AMR3 \\[4pt]
   \hline
    14h 13m & 6h 16m &  1h 30m & 0h 9m & 0h 17m \\ \hline
 \end{tabular}
  \label{table:tracing_runs}

\end{table}

Table~\ref{table:tracing_runs} shows the wall-clock times for the EUHFORIA, Middle$_{EQ}$, Middle$_{STR}$, and Low$_{STR}$ with AMR level 2 and AMR level 3  simulation runs. 
All the simulations are performed on 1 node with 36 compute cores on the Genius cluster at the Vlaams Supercomputing Centre\footnote{\url{https://www.vscentrum.be/}}. 
It is important to remark, first of all, that the Middle$_{EQ}$ simulation with Icarus is significantly faster than the EUHFORIA simulation, even though they use very similar grids and resolutions. The main difference between the two codes is the numerical schemes: Icarus uses a finite volume scheme, while constrained transport is used in EUHFORIA. 
It should also be noted that the AMR level 2 and 3 runs are a factor of ten and 5.29 faster, than the Middle$_{STR}$ simulation, respectively. Yet, they produce very similar results, as noted above. This is because the AMR level 2 and 3 runs start from the basic coarse Low$_{STR}$ grid. The AMR level 2 grid thus has the same resolution as the Middle$_{STR}$ grid but only where the AMR criterion is satisfied (i.e.\ in the CME plasma in this case). For the AMR level 3 simulation, the resolution is locally even twice as high as that in the Middle$_{STR}$ grid, but only where the AMR criterion is satisfied.

\subsection{Shock refinement} \label{section:shock_refinement}
The next AMR criterion that we consider is aimed at refining the CME shock. Shock waves can be located in the simulation domain by looking for regions where $\nabla\cdot\vec{V} < 0$, with \vec{V} denoting the local plasma velocity. 
This approach is used by the EUHFORIA shock tracer introduced in \citet{Wijsen2022}. 
Since this criterion includes all shocks within the computational domain and not only those characteristic to the CME shock front, we introduced additional criteria: we only refined the compressed areas along the Sun-Earth line (in the range of ($lon_{earth} \pm 30^\circ$)), near the equatorial plane (in the range of ($clt_{earth} \pm 10^\circ$) and in the hemisphere of the domain where the CME is injected (in the range of ($lon_{CME} \pm 90^\circ$). From a forecasting perspective, this approach is reasonable. Therefore, we also used it in the next section where we combined the different mesh refinement criteria.

\begin{figure}[hbp!]
    \centering
    \includegraphics[width=0.5\textwidth]{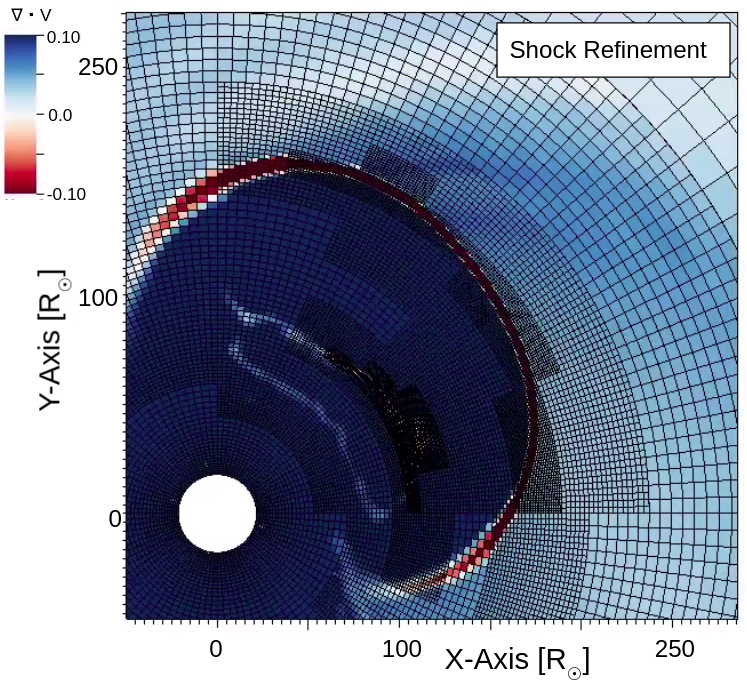}
    \caption{$(\nabla \cdot \mathbf{V})$ criterion from the AMR level 4 simulation run (yielding approximately the same radial resolution at L1 as the EUHFORIA run that is shown in Fig.~\ref{fig:div_arrival_shock_width} for comparison) focusing on the CME shock front. The maximum positive and negative values in the domain are normalised to +0.05 and -0.05, respectively. Blue and red denote areas of expansion and compression, respectively.}
    \label{fig:amr_level_4_shock}
\end{figure}

In Fig.~\ref{fig:amr_level_4_shock} we present an example of the $(\nabla \cdot \mathbf{V})$ criterion in the equatorial plane. 
The shock front of the CME is shown in red in the figure, indicating that the shock front has solely negative $(\nabla \cdot \mathbf{V})$ values and the area that matches the implemented criterion is refined to level 4. It is important to note that AMR up to level 4 yields a slightly lower radial resolution and AMR level 5 yields a slightly higher  one at L1 than the EUHFORIA run that is shown in Table~\ref{table:amr_resolutions_at_l1}.

\begin{figure*}[htp!]
    \centering
    \includegraphics[width=\textwidth]{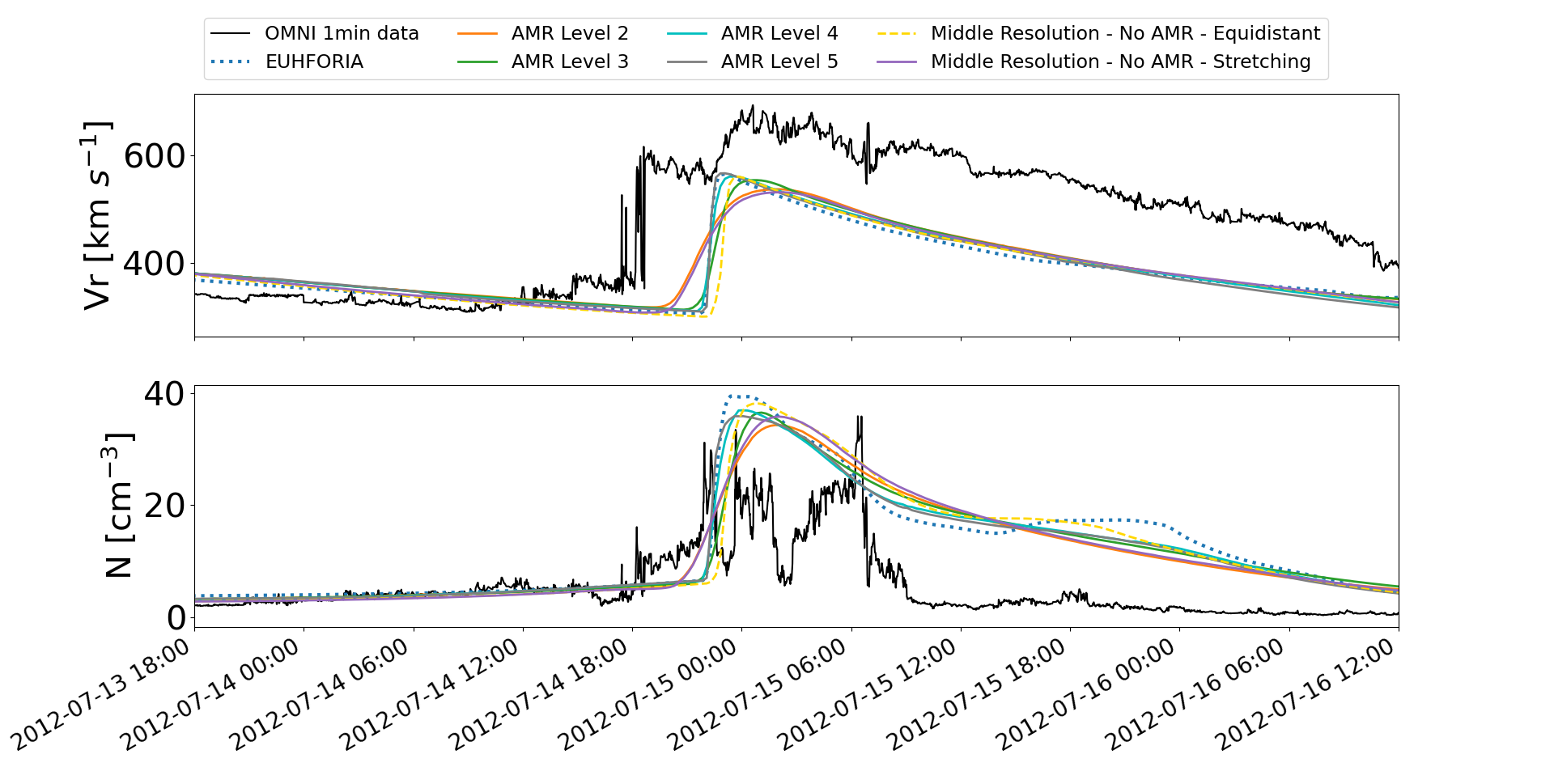}
    \caption{Time series at the Earth for the radial velocity values given in [km s$^{-1}$] and number density given in [cm$^{-3}$]. The divergence of the velocity is used as the AMR criterion. The black curve shows the observed data, the blue one corresponds to the original EUHFORIA result for the high-resolution simulation, and the orange, green, cyan, grey, yellow, and purple curves correspond, respectively, to the AMR level 2, 3, 4 and 5 simulations, the middle-resolution simulation without AMR performed on an equidistant grid without stretching, and the middle-resolution run without AMR with grid stretching applied to the computational domain.  The horizontal axis shows the time evolution.}
    \label{fig:obs_vs_amr_div_euhforia}
\end{figure*}

Figure~\ref{fig:obs_vs_amr_div_euhforia} shows the time series of the different simulations together with the OMNI-1min data. 
The upper panel shows the radial velocity data and the bottom panel shows the number density.
From this figure we can see that the shock jump at the CME arrival at the midnight of July 15 is the sharpest in the data curve representing the EUHFORIA simulation, given by the blue curve, and the AMR level 5 simulation, given by the grey curve. 
Locally at L1, AMR level 2 corresponds to the middle resolution, and that is also what we see in the comparison.
In order to analyse these data, we again consider the same parameters as for the first refinement criterion in Sect.~\ref{section:dynamic_refinement}, that is, the shock arrival time and the shock width. In addition, we look at the maximum peak speed and the average gradient of the shock.

\begin{figure}[hbt!]
    \centering
    \includegraphics[width=0.5\textwidth]{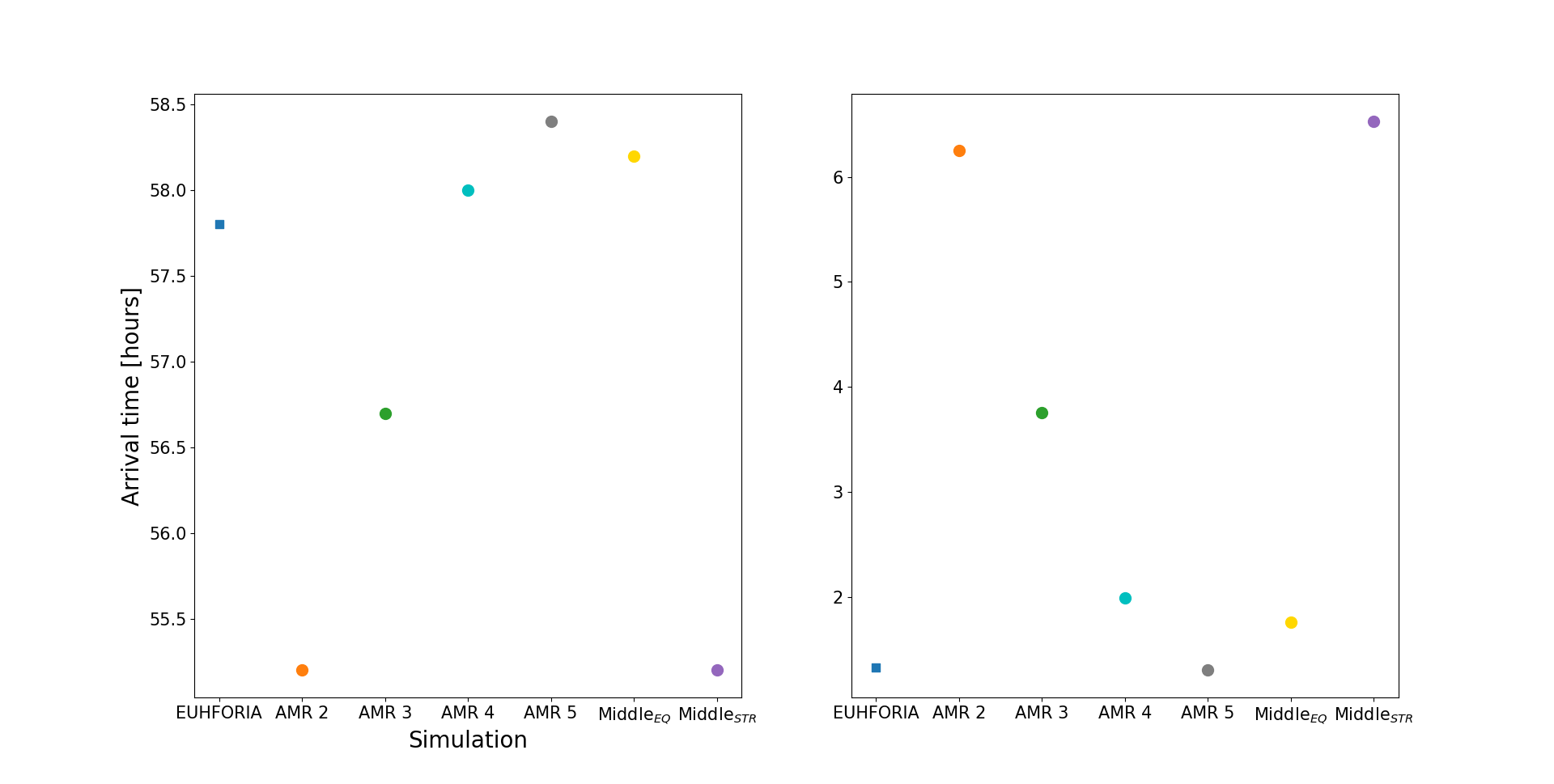}
    \caption{Arrival times (left) and shock widths (right) upon arrival for the original EUHFORIA simulation, the runs with AMR level 2, 3, 4, and 5, and the middle-resolution runs on the equidistant and stretched grids. The horizontal axis indicates the simulation, and the vertical axis shows the arrival times and shock widths in $\;h$.}
    \label{fig:div_arrival_shock_width}
\end{figure}

\begin{figure}[hbt!]
    \centering
    \includegraphics[width=0.5\textwidth]{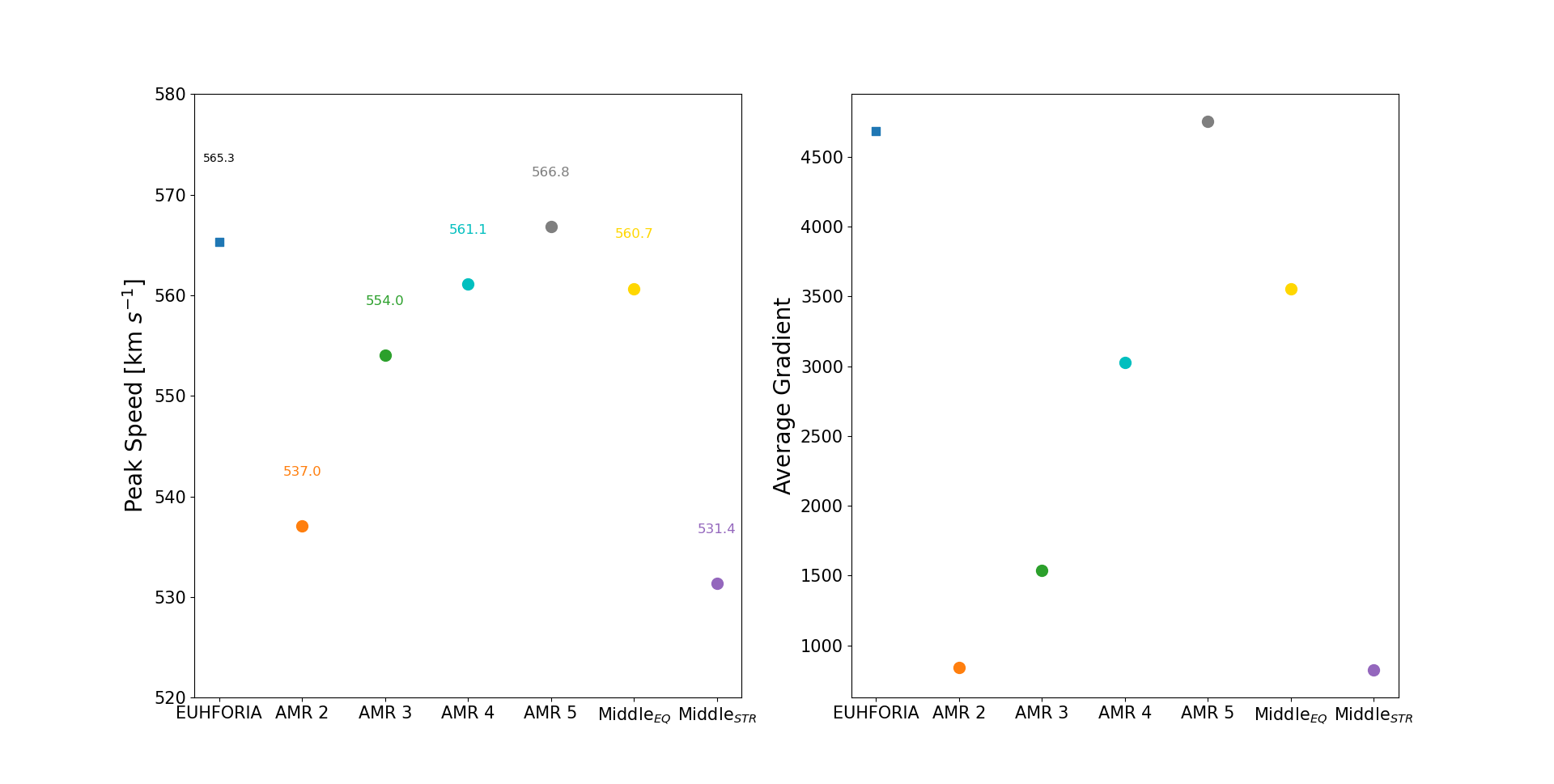}
    \caption{Vertical  axis shows the maximum peak velocities (left) and the average gradients (right) upon arrival for the original EUHFORIA simulation, the runs with AMR levels 2, 3, 4, and 5, and the middle-resolution runs on the equidistant and stretched grids. The horizontal axis indicates the simulation type.}
    \label{fig:div_gradient_speedpeaks}
\end{figure}

Figure~\ref{fig:div_arrival_shock_width} shows the arrival times of the CME at the Earth and the shock width for the different simulations. 
On the left, we see the different arrival times for the seven performed simulations. 
All the parameters for the original EUHFORIA, Icarus, Middle$_{EQ}$, and Middle$_{STR}$ simulations are the same as considered in the previous section, since we used the same simulations results. The results for the simulations with the similar (AMR levels 4 and 5) local resolution at L1 as the EUHFORIA run yield similar values for both the arrival time and the shock widths.
Also, Figs.~\ref{fig:obs_vs_amr_div_euhforia}, \ref{fig:div_arrival_shock_width}, and~\ref{fig:div_gradient_speedpeaks}, show that the simulation with AMR level 2 gives very similar results to the middle-resolution simulation on a stretched grid. The AMR level 3 run arrival time is $56.5$ h, while for the AMR level 4 and 5 simulations, the arrival times are $57.9\;$h and $58.4\;$h, respectively, and for the EUHFORIA simulation, the shock arrives at $57.8\;$h. 
The right panel of Fig.~\ref{fig:div_arrival_shock_width} shows the shock widths for the same simulations as shown in
Fig.~\ref{fig:obs_vs_amr_div_euhforia} with the same corresponding colours. Here we can see that the shocks modelled by the equidistant EUHFORIA and Icarus simulations are similar to the AMR level 4 and 5 runs, and correspond to $1.99\;$h and $1.30\;$h, respectively. 
The run with AMR level 2 and the middle-resolution simulation without AMR on the stretched grid have the largest shock widths, both amounting to $6.7\;$h. 
The shock width for the EUHFORIA simulation is $1.33\;$h, while for Icarus Middle$_{EQ}$, the shock width is $1.76$. 

The arrival times and the shock widths together give information about the shock features. We can see that in the simulations with higher- (local) resolution computational domains, the shocks are narrower, they are more localised, and they arrive a little later, while in the simulations with the lowest resolutions, the shocks are more diffused and smeared out. Therefore, they arrive earlier and the shock widths are larger in these simulations. Thus, the shock widths converge as we increase the AMR levels imposed in the domain, as expected.

 Figure~\ref{fig:div_gradient_speedpeaks} shows the maximum peak speed and the average gradient of the shock for the different simulations. 
On the left-hand side, we see the peak velocity values at L1 obtained with the different simulations. As we can see in Figure~\ref{fig:obs_vs_amr_div_euhforia}, the peak speeds for EUHFORIA, the run with AMR levels 4 and 5, and the Icarus middle-resolution simulation on an equidistant grid resulted in sharper shocks at L1. These values are very close to each other for all simulation runs, except for the AMR level 5 run, which is sharper than the rest of the simulations. However, the orange and purple points, corresponding to the run with AMR level 2 and the middle-resolution simulation on a stretched grid, respectively, have similar and slightly lower peak velocity values. The right-hand side figure shows the average gradients for the shocks for the performed simulations. Average gradients can be used to assess the steepness of the shocks. The shocks obtained with the equidistant EUHFORIA and the Icarus simulations, and the AMR level 5 simulation result in  the highest gradients (i.e.\ they yield the sharpest shocks). The AMR level 4 shock is significantly sharper than the AMR level 3 shock. The least steep shocks are the ones corresponding to the AMR level 2 simulation and the middle-resolution simulation without AMR on the stretched grid, which have the same local resolution at L1 and indeed also yield the same average gradient. The obtained average gradients in the CME shock is in good agreement with the shock widths shown in Fig.~\ref{fig:div_arrival_shock_width} (right-hand side panel). The shock width and the average gradients of the individual simulation then explain the differences in the maximum peak speeds and the arrival times in each case, because the solar wind values before the CME shock arrival are very similar.

Table~\ref{table:times_div_AMR} shows the wall-clock CPU times required for each AMR run with the shock refinement criterion, while the Middle$_{EQ}$, Middle$_{STR}$, and EUHFORIA simulation times are given in Table~\ref{table:tracing_runs} as a reference. Since we substantially limit the regions to refine to those that are compressed (in the CME shock) and only on the Earth side of the Sun, and in the neighbourhood of the equatorial plane, the simulations are very fast, because only small portion of the domain is refined at any moment in the simulation. For the case studied here, the AMR level 2 simulation takes only 8 minutes, the AMR level 3 simulation takes 11 minutes, the AMR level 4 simulation takes 27 minutes, and the AMR level 5 run takes 2h and 39 minutes, which produces the sharpest shocks in the domain. AMR level 4 and 5 simulations are thus 14 and 2.3 times faster than Middle$_{EQ}$ simulation, respectively. Given that the modelled data is comparable to the original equidistant simulations, we can conclude that this refinement criterion speeds up the simulations very significantly. It is important to remember that the grid for the run with AMR level 4 has almost the same resolution at L1 as the grid in the EUHFORIA simulation, and for the AMR level 5 case, the resolution at L1 is higher than in the EUHFORIA run. Yet, the AMR level 4 run with this AMR criterion (only refining the CME shock front) is 31.6 times faster than EUHFORIA, and the AMR level 5 run is 5.4 times faster than EUHFORIA.

\begin{table}[htb!]
 \begin{center}
  \caption{Run times required for the AMR level 2, 3, 4, and 5 simulation runs for the shock refinement criterion on the stretched grid. $\tau_{\text{EUHFORIA}}$ and $\tau_{\text{Icarus}}$ show the speed-up factors of each AMR simulation compared to the EUHFORIA and Icarus middle-resolution simulations given in Table~\ref{table:tracing_runs}. All the simulations were performed on 1 node with 36 processors on the Genius cluster at the Vlaams Supercomputing Centre.}
  \begin{tabular}{c c c c c }
  \hline\hline
   & AMR 2& AMR 3 & AMR 4 & AMR 5  \\[4pt]
   \hline
   Run time & 0h 8m & 0h 11m&  0h 27m  & 2h 39m \\ 
   \hline
   $\tau_{\text{EUHFORIA}}$&  106.6 &  77.5 & 31.6 & 5.4 \\ 
   \hline
    $\tau_{\text{Icarus}}$&  47.0 &  34.2 & 13.9 & 2.3  \\ 
   \hline
 \end{tabular}
 \label{table:times_div_AMR}
\end{center}
\end{table}

\subsection{Combined refinement} \label{section:combined_AMR}

\begin{figure}[hbt!]
    \centering
    \includegraphics[width=0.5\textwidth]{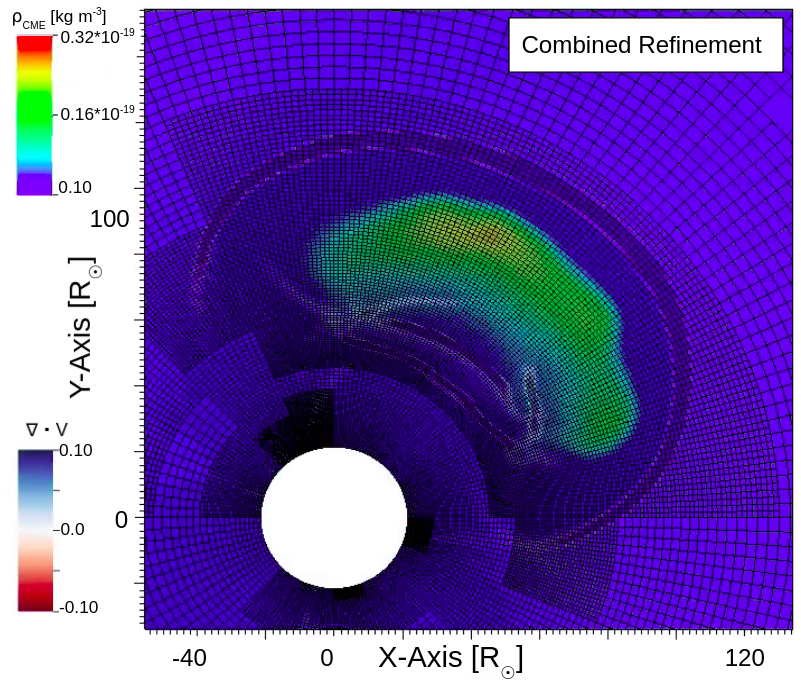}
    \caption{AMR level 3 for the combined tracing function. The $X$- and $Y$-axis show the distance from the centre in solar radii. The CME density tracing function is shown in the equatorial plane with the corresponding colour map.}
    \label{fig:amr_level_3_combi_function}
\end{figure}

Lastly, we consider the combined refinement criterion. In these simulations, we combined the refinement criteria of CME density tracing and shock capturing, limiting the refinement to the Earth-oriented side of the domain and to a narrow latitudinal band around the equatorial plane, as before. Thus, throughout the simulation both the CME itself and the CME shock front are refined. Figure~\ref{fig:amr_level_3_combi_function} shows the refined grid with the combined refinement criteria in the equatorial plane. For better visibility the CME area is zoomed in. In this case, both the CME itself and its shock front are refined to level 3; the rest of the domain is refined to the base refinement level. The colour map given in this figure is a combination of the two different colouring schemes. The shock front is coloured according to the ($\nabla \cdot \text{V}$) values. The maximum positive and negative values in the domain are normalised to +0.1 and -0.1, respectively. We see that both the CME interior and its shock-front are fully covered by the highest refinement level present in the domain, because the refinement condition implemented in the code takes into consideration both the CME cloud and the shocks. 

\begin{figure*}[hbt!]
    \centering
    \includegraphics[width=\textwidth]{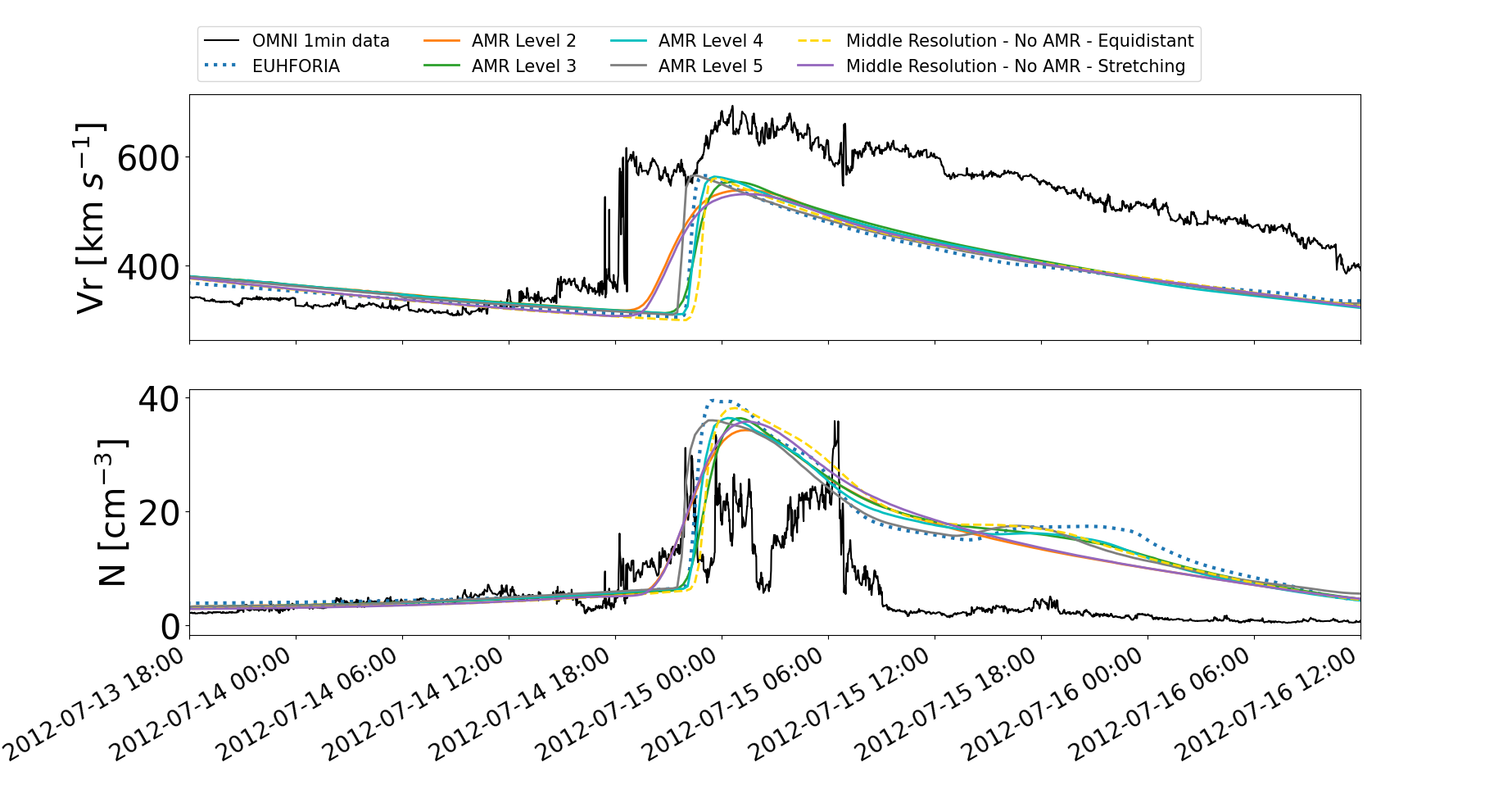}
    \caption{Time series at the Earth for the radial velocity values given in [km s$^{-1}$] and number density given in [cm$^{-3}$]. Both, the CME tracing function and the divergence of $\Vec{V}$ is used for the AMR criteria. The black curve shows the observed values, the blue line corresponds to the original EUHFORIA result from the high-resolution simulation, and the orange, green, cyan, grey, yellow and purple  lines correspond, respectively, to the AMR level 2, 3, 4, and 5 simulations, the middle-resolution simulation without AMR performed on an equidistant grid without stretching, and the middle-resolution run without AMR with grid stretching applied to the computational domain. The horizontal axis shows the time evolution.}
    \label{fig:obs_vs_amr_combi_euhforia}
\end{figure*}

Figure~\ref{fig:obs_vs_amr_combi_euhforia} shows the obtained time-series data at Earth. As in the previous two cases, we show the observed data in black and original equidistant EUHFORIA result in blue. Yellow and purple correspond to the Icarus results without AMR on the middle-resolution domain without and with grid stretching, respectively. The orange line corresponds to the AMR level 2 simulation results, the green line represents the AMR level 3 results, the cyan line corresponds to AMR level 4 run and the grey line corresponds to AMR level 5 simulation. The synthetic AMR level 4 and 5 data are significantly better than the AMR level 2 and level 3 data, but in the AMR level 4 case, the CME shock is not as sharp as in the original simulations using an equidistant grid, where AMR level 5 manages to produce an even sharper shock than in the equidistant simulations. Similar to the previous refinement criterion, the AMR level 2 and middle-resolution simulations on a stretched grid show very similar profiles, as in these simulations the local cell sizes at L1 are the same.

\begin{figure}[hbt!]
    \centering
    \includegraphics[width=0.5\textwidth]{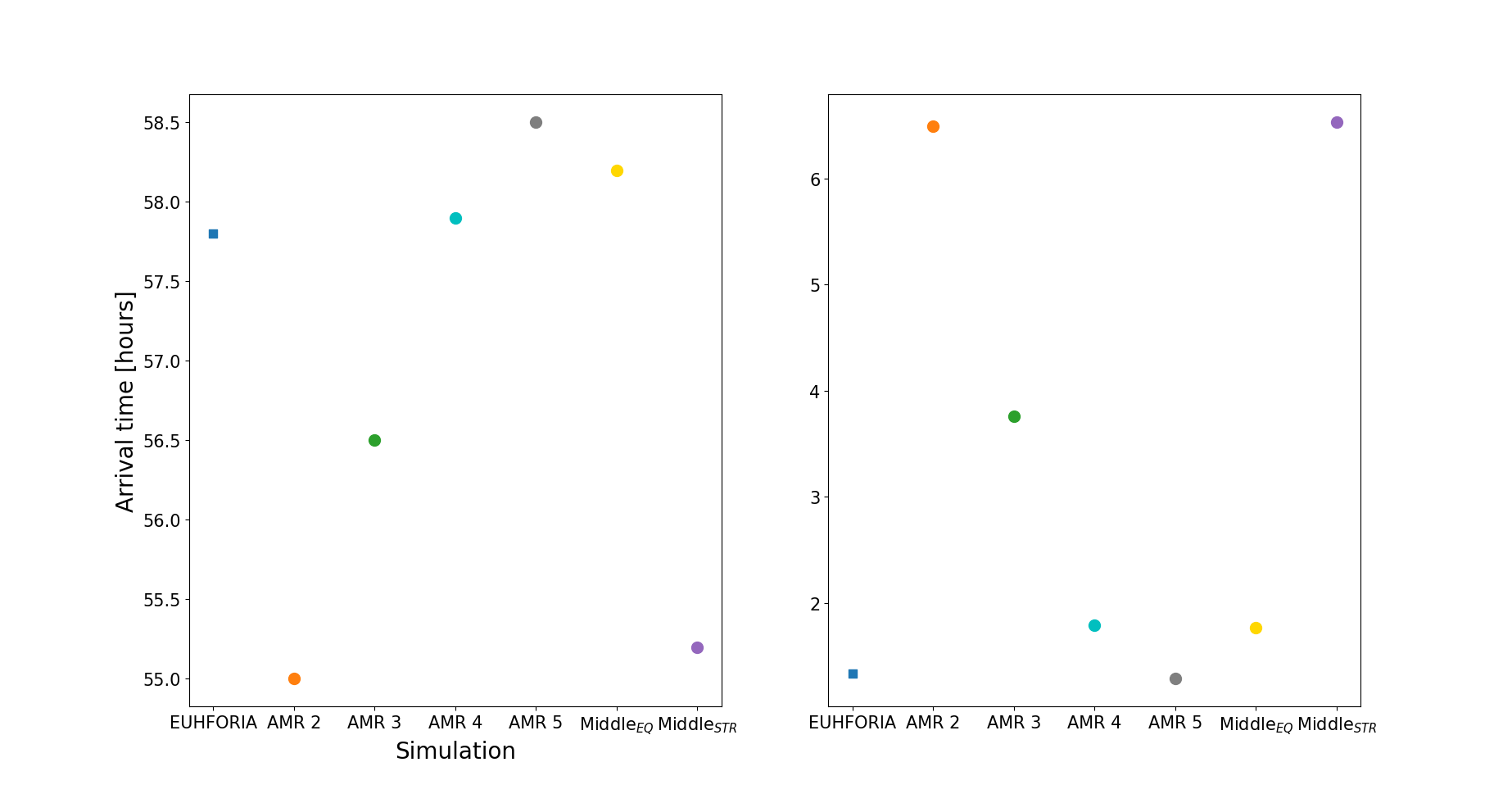}
    \caption{Arrival times (left) and shock widths (right) upon arrival for the original EUHFORIA simulation, the AMR level 2, 3, 4, and 5 runs, and the middle-resolution runs on the equidistant and stretched grids, respectively. The horizontal axis indicates the simulation type and the vertical axes show the arrival times (left) and the shock widths in$\;h$ (right). }
    \label{fig:combi_arrival_shock_width}
\end{figure}

\begin{table}[h!]
 \begin{center}
  \caption{Run wall-clock times required for different refinement level runs for the combined criterion on the radially stretched grid. $\tau_{\text{EUHFORIA}}$ and $\tau_{\text{Icarus}}$ show the speed-up factors of each AMR simulation compared to the EUHFORIA and Icarus middle-resolution simulations given in Table~\ref{table:tracing_runs}. All the simulations were performed on 1 node with 36 processors on the Genius cluster at the Vlaams Supercomputing Centre.}
  \begin{tabular}{c c c c c}
  \hline\hline
  &AMR2& AMR3& AMR4 & AMR5 \\[4pt]
   \hline
 Run time & 0h 9m  & 0h 14m & 0h 40m & 3h 59m \\ \hline
   $\tau_{\text{EUHFORIA}}$&  94.8 & 60.9 & 21.3 & 3.8  \\ 
   \hline
    $\tau_{\text{Icarus}}$& 41.8 & 26.9 & 9.4  & 1.57\\ 
   \hline
 \end{tabular}
 \label{table:times_combi_AMR}
\end{center}
\end{table}

In order to analyse the synthetic time-series data, we address the same parameters as in the previous subsections.  Figure~\ref{fig:combi_arrival_shock_width} shows the shock arrival times obtained with the different simulations (left) and the shock widths upon the arrival of the CME at the Earth (right). The simulations using an equidistant grid yield shocks that are arriving about 3$\;$h later and these simulations also yield the narrowest shock widths upon the arrival. In the AMR level 2 and Middle$_{STR}$ runs (with the same radial grid size at L1) the CME shocks arrive about 3$\;$h earlier and these low-resolution simulations yield the widest shock widths. In the AMR level 3 simulation, the shock arrives about $1.5\;$h later but still also about $1.5\;$h earlier than in the simulations with an equidistant grid. For this simulation, the width of the CME shock is also in between the shock widths obtained with lower-resolution stretched grids and (higher-resolution) equidistant grids. For the AMR level 4 simulation, however, the shock arrives more or less at the same time as in the simulations with an equidistant grid. The shock width for the AMR level 4 simulation is $1.79\;h$, and for the AMR level 5 simulation it is $1.28\;h$. The results modelled with AMR level 5 are the most accurate. We saw the same tendency with the previous refinement criterion, which was targeting the shock in the simulations. It is worth noting that in this analysis, we assess the shock features and the arrival times for each simulation, since the cone model does not have a complex internal structure. Therefore, it is expected that the features of the shock will be mostly affected by the refinement of the shock parts and the CME refinement will not improve the results significantly, as shown in Sect.~\ref{section:dynamic_refinement}. 

Table~\ref{table:times_combi_AMR} shows the times for the AMR level 2, 3, and 4 simulations for the combined refinement criterion. For the combined AMR criteria, the mesh resolution is refined in a larger area than in the two criteria separately. However, it should be remembered that here we also limited the mesh refinement region to the Earth-side of the Sun and to a narrow latitudinal bandwidth around the equator. This explains why the AMR level 2 and 3 simulations took  1~min and 4~min longer, respectively, than in the previous section (due to the additional refinement of the CME cloud), but the same time (for level 2) or even 3~minutes less (for level 3) than the simulation with only the density tracing criterion (which was not restricted to only the Earth side of the heliosphere and not limited to a narrow bandwidth around the equator). The AMR level 4 simulation took 40~min, almost twice as long as with the simulation with only the shock refinement criterion.  
Hence, the AMR level 4 simulation run was 9.4 times shorter than the Middle$_{EQ}$ simulation in Icarus and 21.3 times shorter than the middle-resolution simulation in EUHFORIA, while the local resolution at L1 is almost the same in these simulations. The AMR level 5 simulation, while producing the most accurate results, was 3.8 times faster than the EUHFORIA simulation and 1.57 times faster than the Icarus simulation with an equidistant grid.

\section{Discussion and Conclusion} \label{section:discussion_conclusion}
In this study we investigated the potential of radial grid stretching in combination with solution AMR in the Icarus solar wind and CME evolution model to improve the results and speed up the simulations. The CME event of July 12, 2012 was modelled with a cone CME. The synthetic data at L1 from these simulations have been given together with the in situ measurements and the synthetic data obtained with the original EUHFORIA model. Icarus, similar to EUHFORIA, models the heliosphere from 21.5$\;R_\odot$ onwards. Hence, these models do not capture the evolution of the CME through the corona, and therefore deal with a number of uncertainties, which can produce large discrepancies between the modelled and the observational data. We analysed several features characterising the shocks and compared them to assess the performance of each simulation. Since in this study a simple cone CME model was used, we only focused on the shock characteristics and the arrival time of the CME. We did not model the magnetic cloud of the CME, which does not allow us to study the CME interior evolution in the heliosphere accurately. As mentioned at the beginning, the goal of this paper is to demonstrate the capabilities of the Icarus model and its flexibility. 

Earlier, \cite{Verbeke2022} considered the radial grid stretching and AMR both separately. They provided an important speed up of the simulations, but grid stretching reduced the radial resolution of the grid locally near the Earth, and therefore decreased the sharpness of the different variable profiles. In order to minimise this effect, we have explored the combination of grid stretching with AMR and investigated the combined effect on the simulations. Three different refinement criteria were considered: CME tracing, shock refinement, and, finally, the combined refinement criterion.

The first criterion considered here is the refinement of the CME described in Sect. \ref{section:dynamic_refinement}. In this case, we did not take into consideration the CME shock front and once the CME expanded, the shock front was not covered by the refined blocks in the domain. We could observe that AMR levels did not have a notable effect on the simulation. In particular, compared to the equidistant Icarus and EUHFORIA results, the heliosphere modelled with a combination of grid stretching and different AMR level simulations did not produce as sharp shocks upon the arrival as desired. As mentioned earlier, the interior of the cone CME is homogeneous and it does not model the internal magnetic field of the CME. Because of this, the refinement of the CME did not improve the results significantly, since there is no structure present within the CME to refine. The CME tracing function can be an interesting and crucial criterion in the cases where the CME model is more complex, such as in the magnetised CME models (e.g. a spheromak or a flux-rope model), because their interior is quite complex and requires high resolution to model the internal magnetic structure accurately.

Following the CME refinement, we investigated the effect of the shock refinement given in Sect. \ref{section:shock_refinement}. This criterion is rather general and can be used with different purposes: to refine the shocks of the solar wind and the shocks associated with the CME or CIRs. In this case, we restricted the refinement along the Sun-Earth line to refine only the propagating CME, because we are interested in data at L1. This approach is very effective from a forecasting point of view, yet it is very flexible and the user can modify the range in the domain freely. The simulations are even faster, since only the portion of the CME shock-front is refined until it reaches the Earth. As a result we obtained high resolution in the domain only where and when necessary.
In order to make the resolution of the shock at L1 near Earth comparable to the equidistant simulations, we investigated the simulations with AMR levels 4 and 5 as well, unlike the first refinement criterion, where we saw that AMR did not have effect on the quality of the simulations.  
In the simulations with AMR levels 4 and 5, the shocks upon the arrival of the CME at the Earth are much sharper than in the previous refinement cases. AMR level 5 is sharper than the shock modelled by the original equidistant grid simulations of Icarus and EUHFORIA. In these cases the CME arrives at the Earth at nearly the same time as in the original EUHFORIA simulation. When comparing to Middle$_{EQ}$ runs in Icarus and EUHFORIA, we observe that AMR level 5 produces better and faster results.

The data modelled by the different AMR simulations converge. AMR level 4 is faster than AMR level 5, but AMR level 5 improves the shock prediction significantly. We can conclude that using AMR level 5 is the optimal choice in this case, since the shock modelled in this run was the best, and the simulation was still faster than the equidistant middle-resolution runs in Icarus and EUHFORIA. Thus, the refinement of the shock in the domain yields good results with these advanced techniques compared to the equidistant and uniformly refined computational domains.

Lastly, the combination of the CME interior and the CME shock front refinement is considered in Sect. \ref{section:combined_AMR}. The main contribution is expected from the shock-front refinement criterion. 
As mentioned earlier, the refinement of the CME interior does not improve the simulations significantly in the cone CME model case, and therefore it was expected that the simulations with the combined refinement and the shock refinement would yield similar results when describing the CME shock-front. The combined criterion gives a rough estimate for the timings of the simulations with complex internal magnetic structure, where both shock-front and the interior must be refined.

After considering the three different refinement criterion for the CME, we can see that the shock refinement and the combined refinement criterion yielded better results in the time-series profiles at the Earth than the density tracing refinement criterion. However, in the latter case, the contribution was mainly from the shock refinement criterion.
The times required by the AMR simulations are significantly shorter than the original equidistant simulations. Comparing the last two criteria, we can see that the shock refinement criterion yielded faster simulations than the combined criterion simulations, yet the results were very similar because the tracing criterion did not improve the shock features. The longer time required by the combined criterion is natural, since additional portion is refined in the domain together with the shock area. Thus, we can conclude that the simulations with the shock front refinement criterion produced the best results with the shortest simulation times, which is crucial for forecasting purposes. In particular, the most optimal combination for modelling the cone CME is the AMR level 5 for the shock front refinement criterion, because it produces the best results faster than the equidistant Icarus and EUHFORIA simulations.

The different AMR criteria presented in this work can be more beneficial for more complex and sophisticated CME models.  Therefore, depending on the simulation setup, different AMR criteria can be used to obtain the best results in the shortest time. In future work, we will consider the implementation of more realistic CME models with a complex internal magnetic field structure, such as the Linear Force-Free Spheromak  \citep{Verbeke2019} and the Gibson\&Low \citep{Gibson1998} models. The density tracing criterion might improve the results significantly for these complex CME models, since it will be able to better resolve the internal magnetic field structures, and hence its evolution.  Such simulations are usually slower than the simple cone CME model simulations, but we will use the advanced techniques explained in this paper to make the simulations as efficient as possible. In some complex cases, such as CME-CME or CME-CIR interactions, a sufficiently high resolution is imperative to capture the correct underlying plasma dynamics. In these scenarios, it would be very computationally expensive to refine the whole domain to the required resolution. Therefore, AMR is a pivotal asset when high resolution is required only in certain regions of the domain. Moreover, the flexibility of the AMR criteria gives large freedom to the user to modify the grid accordingly, and thus obtain better and faster results without needing to spend excessive computational resources.

\begin{acknowledgements}
This research has received funding from the European Union’s Horizon 2020 research and innovation programme under grant agreement No 870405 (EUHFORIA 2.0) and the ESA project "Heliospheric modelling techniques“ (Contract No. 4000133080/20/NL/CRS).
These results were also obtained in the framework of the projects C14/19/089  (C1 project Internal Funds KU Leuven), G.0D07.19N  (FWO-Vlaanderen), SIDC Data Exploitation (ESA Prodex-12), and Belspo project B2/191/P1/SWiM.
The Computational resources and services used in this work were provided by the VSC-Flemish Supercomputer Center, funded by the Research Foundation Flanders (FWO) and the Flemish Government-Department EWI.
The authors thank Dr.\ Emmanuel Chan\'e for many interesting discussions and his very valuable input on this paper.
\end{acknowledgements}


\bibliographystyle{aa}
\bibliography{bibliography}
\end{document}